# Microstructured large-area photoconductive terahertz emitters driven at high average power

**Mohsen Khalili,**[1,*] **Tim Vogel,**[1] **Yicheng Wang,**[1] **Samira Mansourzadeh,**[1] **Abhishek Singh**[2,3], **Stephan Winnerl**[2], **and Clara J. Saraceno**[1]

[1]*Photonics and Ultrafast Laser Science, Ruhr-University Bochum,44801, Germany*
[2]*Helmholtz-Zentrum Dresden-Rossendorf, Institute of Ion Beam Physics and Materials Research, 01328 Dresden, Germany*
[3]*Centre for Advanced Electronics, Indian Institute of Technology Indore, Indore, India*

*Mohsen.Khalilikelaki@ruhr-uni-bochum.de

**Abstract:** Emitters based on photoconductive materials excited by ultrafast lasers are well-established and popular devices for THz generation. However, so far, these emitters – both photoconductive antennas and large area emitters - were mostly explored using driving lasers with moderate average powers (either fiber lasers with up to hundreds of milliwatts or Ti:Sapphire systems up to few watts). In this paper, we explore the use of high-power, MHz repetition rate Ytterbium (Yb) based oscillator for THz emission using a microstructured large-area photoconductive emitter, consist of semi-insulating GaAs with a 10×10 mm² active area. As a driving source, we use a frequency-doubled home-built high average power ultrafast Yb-oscillator, delivering 22 W of average power, 115 fs pulses with 91 MHz repetition rate at a central wavelength of 516 nm. When applying 9 W of average power (after an optical chopper with a duty cycle of 50%) on the structure without optimized heatsinking, we obtain 65 µW THz average power, 4 THz bandwidth; furthermore, we safely apply up to 18 W of power on the structure without observing damage. We investigate the impact of excitation power, bias voltage, optical fluence, and their interplay on the emitter performance and explore in detail the sources of thermal load originating from electrical and optical power. Optical power is found to have a more critical impact on large area photoconductive emitter saturation than electrical power, thus optimized heatsinking will allow us to improve the conversion efficiency in the near future towards much higher emitter power. This work paves the way towards achieving hundreds of MHz or even GHz repetition rates, high-power THz sources based on photoconductive emitters, that are of great interest for example for future THz imaging applications.

## 1. Introduction

THz time-domain spectroscopy based on ultrafast near-infrared (NIR) laser drivers is widely applied in time-resolved molecular spectroscopy, imaging, and many more applications in both science and technology [1–4]. Most of these applications could benefit from higher THz average powers to increase signal-to-noise ratio and shorten measurement times, thus a recent area of research has emerged focusing on applying modern high-power lasers for THz generation, nowadays reaching close watt-level sources [5,6]. In these recent demonstrations, mostly optical nonlinear methods like optical rectification or two-color plasma filaments were explored. These are attractive due to the wide availability of nonlinear crystals and the easiness of using gases or air in the case of plasma sources. However, THz conversion efficiency in these methods strongly relies on high driving pulse energies [7–9]. Typically, minimum pulse energies of tens to hundreds of microjoules or higher are imperative for efficient conversion [7,10]. Therefore, unless prohibitively high average powers are available, this limits the conversion efficiency for systems operating with repetition rates significantly beyond 1 MHz, which can clearly be seen in the current state-of-the-art in Fig. 1.

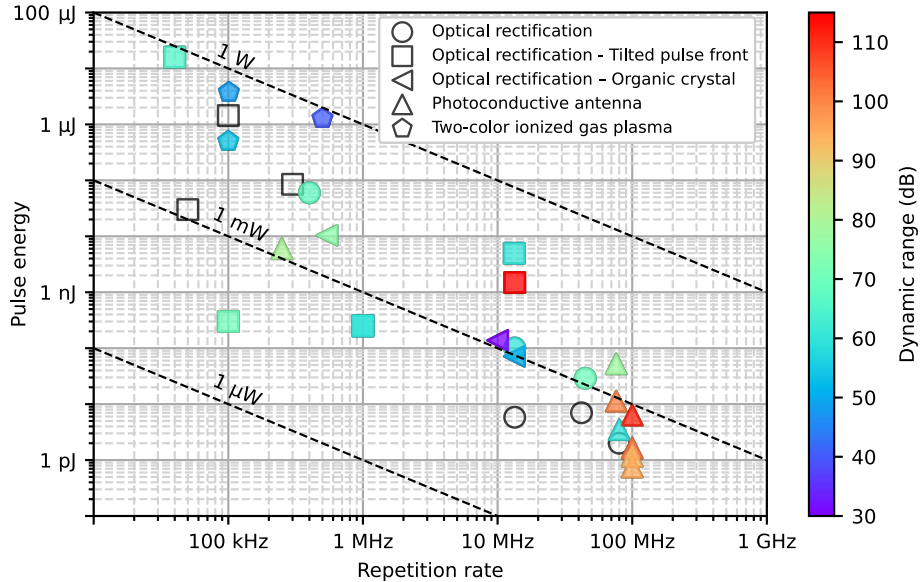

Fig. 1. THz pulse energy versus excitation pulse repetition rate for different THz generation methods. The color bar indicates the peak dynamic range in frequency domain of THz sources. Empty symbols, the dynamic range are not reported.

Extending these current achievements to reach the watt-level and beyond at high repetition rates >>1MHz would enable for example THz imaging at higher frame rate, reduced measurement times and possibly even real-time imaging on the long-run. In this regard, photoconductive (PC) emitter are an excellent alternative [11,12]. The basic principle of THz emission in PC emitter is based on charge carriers which are generated in the photoconductive material by an ultrafast laser pulse and accelerated via an external bias field. As such, much of the performance of PC emitters are intrinsic to the accelerating static field, thus relaxing the requirements on the excitation pulses. In particular, since the energy for acceleration stems from the bias voltage supply, the efficiency is not determined by the Manley–Rowe limit where one near-infrared photon is required for a THz photon. There are two families of PC emitters: photoconductive antennas (PCAs) and large area emitters (LAEs). PCAs consist of a high-resistivity semiconductor substrate with two electrodes constructed on one side of the substrate. The electrodes are connected to an external bias voltage to induce the acceleration field for photoinduced carriers. The excitation power of conventional PCAs are limited to less than 100 mW [13,14], mainly due to the typical dimension of a few µm$^2$ for the photoconductive gap. Therefore, these structures are not well suited for Yb state-of-the-art lasers with hundreds of watts of average power [15,16]. PCAs provide high efficiency for NIR to THz conversion at moderate pulse energies in the nJ range and can be modulated at extremely high frequencies, thus offering the possibility of very high dynamic range (DR). State-of-the-art PCAs exhibit high DR of 111 dB with an emitter THz power of 637 µW, generated with 28 mW of the excitation power, corresponding to 3.4% of conversion efficiency [13,17]. In addition to these results, PCAs with special plasmonic structures have demonstrated 4 mW of THz power with 720 mW of excitation power [18]. In addition to these results, PCAs with special plasmonic structures have demonstrated 4 mW of THz power with 720 mW of excitation power [18]. A new design of PCAs is demonstrated in [19] which is called photoconductive connected array (PCCA) to generate mw-level THz power with maximum 500 mW laser input power. In this antenna geometry, an $N{\times}N$ array of connected PCAs that can cover a larger area of PC without a necessity of a tight focus of the laser beam. An array of microlens is used on top layer to focus the laser beam directly to PC gap, in order to enhance the coupling efficiency of incident laser beam to the active area. However, the maximum incident power was limited to 500 mW. With 400 V bias voltage, a THz power of 712 µW is reported.

An increase in the emitter area is in principle possible with PCAs, with millimeter-sized electrode separations that could handle higher optical power, but these require bias voltage in kV range [14,20]. This problem can be circumvented with microstructured inter-digitated large-area photoconductive THz emitters (LAE) [21–24], operating on the same principles as PCAs but their large active area in the range of 1-100 mm$^2$ allows for the use of higher average and peak power ultrafast driving lasers and moderate bias voltages in the order of a few volts. Furthermore, such emitters can be designed for generation of THz vector beams and for polarization switching [25,26]. So far, LAEs have mostly been explored with ultrafast laser sources based on Ti:Sapphire; both modelocked oscillators with 10 nJ pulse energy at 78 MHz or amplifier based with higher energy of few µJ at 250 kHz, leading to conversion efficiencies of $2.5 \times 10^{-4}$ and $2 \times 10^{-3}$, respectively [21,23]. While these 10 mm²-sized LAEs have been explored for scaling fluence and achieving high fields [23], excitation of LAEs in high average power and high repetition rate regime is still untapped, as a potential to bring average power and DR of THz sources even higher than current records [13,18,23]. Some important parameters of the discussed photoconductive emitters are summarized in the Table .1.

Table 1. Comparison different photoconductive emitters performances.

| Generation method | Optical power | Repetition rate | THz power | Conversion efficiency | Dynamic range | Measurement time |
|---|---|---|---|---|---|---|
| PCA [13] | 28 mW | 100 MHz | 637 µW | 3.4% | 111 dB | 120 s |
| Plasmonic PCA [18] | 720 mW | 76 MHz | 4 mW | 0.55% | 70 dB | Not reported |
| PCCA [19] | 500 mW | 80 MHz | 712 µW | 0.14% | 50 dB | Not reported |
| LAE [23] | 800 mW | 250 kHz | 1.5 mW | 0.2% | 80 dB | 20 s |

In this paper, we explore THz emission from a semi-insulating GaAs based LAE with a 10×10 mm$^2$ active area, within the MHz repetition rate and high average power regime. Our LAE is excited by a frequency-doubled, home-built, high average power ultrafast oscillator, capable of delivering 115 fs pulses at a central frequency of 516 nm, operating at a 91 MHz repetition rate, and offering up to 22 W of average power. The behavior of the LAE was investigated for various excitation spot sizes, optical and electrical duty cycles to understand current limiting factors; in particular, we found that the optical power is the main source of efficiency limitation pointing to the next step by optimizing heatsinking. In fact, in our experiment, the best result was obtained with 9 W of pump power impingent on the structure (after an optical chopper with duty cycle of 50%), reaching 65 µW of THz average power and spectral bandwidth up to 4 THz. Larger powers could not be applied by maintaining the efficiency, due to thermal limitations. Remarkably the LAE could nevertheless be operated without signs of degradation or damage at 18 W of incident average power. We believe our study represents the first important step towards high-average power THz sources at ultra-high repetition rates using LAEs.

## 2. Experimental setup

Our driving laser source, as depicted in Fig. 2, is a home-built Kerr-lens modelocked (KLM) thin-disk laser (TDL) with a separate gain medium (Yb:garnet disk) and Kerr medium (3 mm YAG). The Yb:garnet disk has a 100 µm thickness with a 10% doping concentration and radius of curvature (RoC) of 2 m. The gain medium is pumped with a multimode fiber-coupled diode at 969 nm, featuring a pump spot diameter of 2.5 mm, with 36 passes to compensate for the low absorption of the pump in the very thin gain medium.

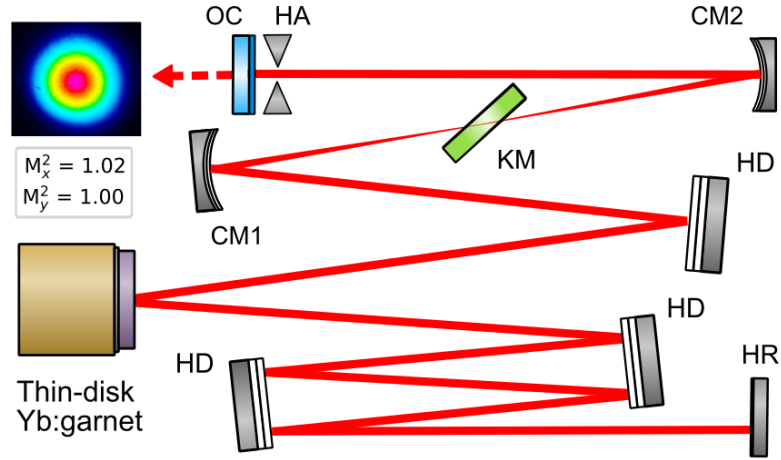

Fig. 2. Kerr-lens modelocked thin-disk laser. HR: high-reflective mirror. CM1 = CM2: high-reflective curved mirrors with RoC = 400 mm. HD: highly-dispersive mirror. HA: hard aperture. OC: output coupler. Inset shows output beam profile and its corresponding beam quality $M^2$ value.

Two identical curved mirrors (RoC= 400 mm) focus the laser mode into the Kerr medium to induce a strong Kerr lens. To stabilize and maintain modelocking, a 1.9 mm pinhole is placed near the output mirror for self-amplitude modulation. The entire oscillator operates in ambient air within a dust-proof enclosure. On days with high humidity (> 60%), the system is purged with dry nitrogen to avoid instabilities. In order to achieve soliton modelocking [27], we balance intracavity self-phase modulation (SPM), primarily arising from the Kerr medium and ambient air, using highly dispersive mirrors with a total roundtrip group delay dispersion (GDD) of -8000 $fs^2$. The single transverse mode ($M^2$=1.02×1.0) oscillator delivers 34 W of average power with a 7% output coupling coefficient at 330 W of pump power. The transform-limited pulse (time bandwidth product, TBP=0.36) with pulse duration of 160 fs (Fig. 3(a)) is centered at 1032 nm with 8 nm bandwidth (Fig. 3(b)). A radio frequency (RF) spectrum with 1 GHz span (Fig. 3(c)) indicates 91 MHz repetition rate and single-pulse operation without any Q-switching instabilities.

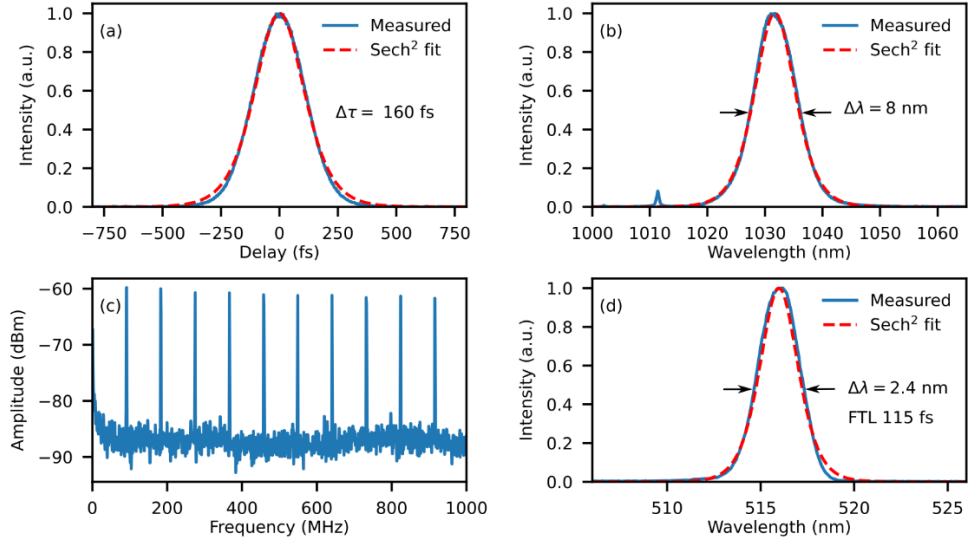

Fig. 3. Characteristic of KLM TDL and its second harmonic. (a) autocorrelation trace. (b) optical spectrum. (c) radio frequency spectrum. (d) optical spectrum of frequency doubled, FTL: Fourier transform limited.

The LAE used for our experiment is a semi-insulating GaAs (SI-GaAs) based emitter with an interdigitated electrode metal-semiconductor-metal (MSM) structure, which is more detailed in [14,22]. The MSM structure is masked by a second metallization layer isolated from the MSM electrodes. The isolation layer on MSM blocks the excitation laser beam in every second period of the MSM structure. In this way, charge carriers are excited and accelerated unidirectionally and radiated THz fields interferes constructively in the far field. Fig. 4 shows the geometry of microstructure LAE. The emitter has a large active area of $10 \times 10$ mm$^2$ which is housed in a water-cooled copper mount that extracts a small fraction of the accumulated heat from the emitter sides. The active area of the emitter composed of interdigitated finger electrodes are created via optical lithography on a SI-GaAs substrate. The electrode structure, with a width and spacing of 10 μm, can produce a bias field on the order of kV/cm at a moderate bias voltage in the range of tens of volts [14].

The experimental setup for THz generation and detection is depicted in Fig. 5. A small fraction of the laser power is separated by a 1% output coupler into the probe arm. A half-waveplate in combination with thin-film-polarizer is used to control the excitation pump power. The photon energy of our oscillator at 1032 nm is smaller than the bandgap and thus our laser is not directly suitable for excitation of the LAE. Therefore, we frequency-double the pump beam using a 2 mm LBO crystal, to operate the LAE at 516 nm and enabling efficient absorption in GaAs. 22 W of green light is generated - approximately 65% second harmonic generation (SHG) efficiency - at the maximum pump power of 34 W. The excitation green pulse is centered at 516 nm, with a fitted full width at half-maximum spectral bandwidth of 2.4 nm (FWHM, $\Delta\lambda$), as shown in Fig. 3(d). This corresponds to a Fourier-transform-limited pulse of 115 fs (FWHM, $\Delta\tau$). The pulse duration of the green beam is in agreement with the compression of the fundamental pulse as a consequence of SHG, $\tau(2\omega) = \tau(\omega)/\sqrt{2}$ [28]. The divergent green beam is collimated to 10 mm in diameter using lens L$_2$ before being focused by a short-focal-length lens (L$_3$ = 50 mm).

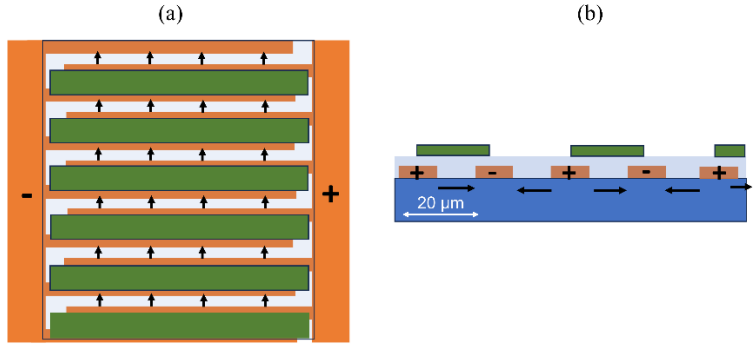

Fig. 4. Geometry of a microstructured LAE. (a) top view of LAE. Electrodes patterns are indicated with orange color, the arrows show the bias field direction. The green pattern is the second metallization to block the incident laser beam, as described in the text. (c) a cross section of the LAE is depicted.

The LAE is placed at normal incidence after the focus of the lens to keep a distance from the tightly focused pump beam and facilitate efficient collection of the highly divergent high-frequency THz components by an off-axis parabolic (OAP) mirror. The generated THz beam maintains the phase front of the incident pump beam [23,29]. The generated THz pulse is collimated and focused into a detection crystal consisting of a 1 mm gallium phosphide (GaP) crystal, using a pair of 50.8 mm diameter gold-coated OAP mirrors with focal lengths of 101.6 mm and 50.8 mm, respectively. The electric field of the emitted THz pulse is characterized using conventional electro-optic sampling (EOS) [30] in combination with standard balanced detection scheme [31]. We note that in future realizations, once the output power is optimized with ideal LAEs and heatsinking, a PC receiver will also be an excellent alternative for detection. The probe beam is periodically delayed relative to the THz beam by a fast-oscillating delay line, set to a frequency of 1 Hz and total scan range of 15 ps. A waveform generator in combination with a voltage amplifier is used to provide a flexible voltage source with different waveshape and modulation frequency to use for bias voltage of the LAE. The voltage signal is simultaneously used to synchronize the lock-in amplifier for low-noise detection. Although it is possible to electronically modulate the signal at MHz frequencies [22], we chose a lower frequency about 997 Hz to maintain consistency with our later experiment when an optical chopper, with a maximum frequency of 1 kHz, is used to modulate the excitation beam while the bias voltage remained constant DC.

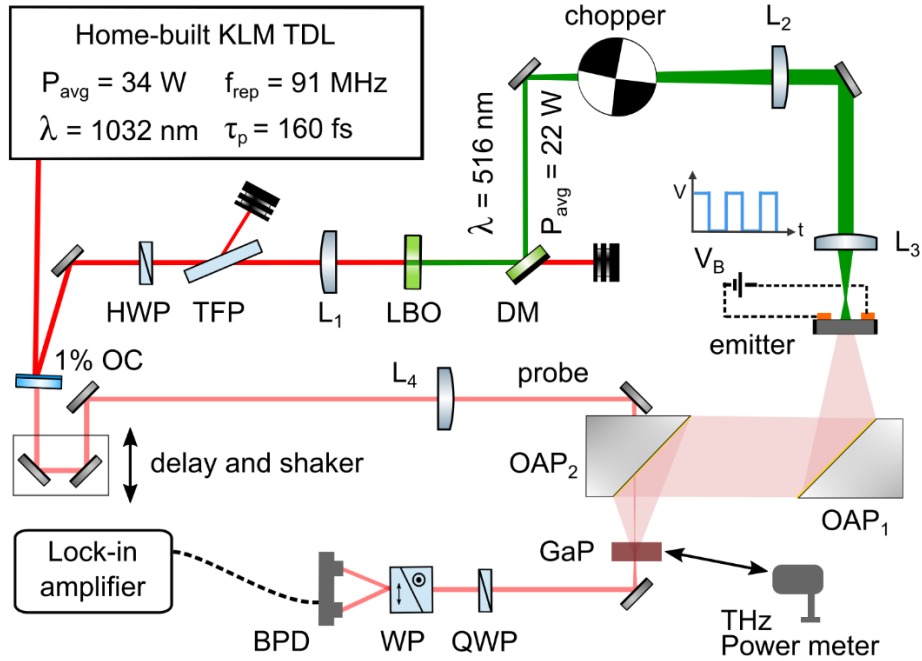

Fig. 5. Schematic of the experimental setup for LAE excitation and THz detection. A 1% output coupler divides the optical pulse trains from the oscillator to pump and probe beams. Lens $L_1$ focuses the laser beam into a 2 mm LBO crystal for second harmonic generation. Lenses ($L_2$, $L_3$) enlarge and collimate the green beam and $L_3$ focuses the beam on the emitter. L4 focuses the probe beam on the detection crystal. HWP: half-waveplate, TFP: thin-film polarizer. GaP: 1 mm gallium phosphide. BPD: balanced photodetector. QWP: quarter-waveplate. WP: Wollaston prism.

## 3.  Results and discussion

The experimental results are presented in the following manner. In the first section, we explore the dependence of the emitted THz field-amplitude and THz average-power on the optical excitation power at three different spot sizes on the LAE. Then using an optical chopper with 50% duty cycle the excitation beam is modulated, effectively reducing the average power while keeping the single pulse energy unchanged. The influence of the optical power and electrical power load on the LAE performance are investigated individually to find which of them has a more critical impact on the LAE saturation.

All experimental results presented in this work are obtained using the same LAE described earlier. The electric field of emitted THz pulses is characterized by EOS as described in the previous section. The THz average power is measured using a pyroelectric thin-film detector (THz 20, SLT GmbH) calibrated by the German metrology institute (PTB). To confirm that the measured THz power is not stray light or heating, we perform background measurements while the pump beam irradiated the LAE with the bias voltage switched off.

### 3.1 LAE saturation behavior: bias modulation, constant optical excitation

We investigate the influence of the fluence on the emitted THz field by varying the excitation spot size on the LAE. By simply adjusting the position of the LAE in the direction of beam propagation, we could control the beam size on the emitter. The corresponding saturation curves are presented in Fig. 6(a - c), where the excitation beam size ranges from 0.6 mm to 1.7 mm in beam waist diameter (intensity $1/e^2$). For measuring the beam size, we used a CMOS (Dataray-BladeCam-HR) camera and fit a Gaussian function to the beam profile. In Fig. 6(a, b) the data corresponding to the peak-to-peak value of the EOS signal in the time domain, with the y-axis representing the voltage from the balanced photodetector, which is proportional to the THz electric field. For all data points, the bias voltage is the same as 10 V peak-to-peak ($V_{pp}$) with a unipolar square waveshape and a 50% duty cycle at 997 Hz. In Fig. 6(c, d) the measured THz power is plotted versus the pump power. In this case, the bias voltage is modulated at 18 Hz, since the power meter is calibrated at this frequency and is limited in response time. The data from both (EOS method and THz power measurement) exhibit a consistent trend for each beam diameter, in particular the fluence values, where maxima in field amplitude and power are reached, coincide. This allows us to rely on EOS measurements for subsequent measurements. Note, however, that the measured THz power does not simply scale quadratically with the amplitude obtained in the EOS measurements. The reason is that for the larger beam waist diameters the THz spot on the EOS crystal is also larger and thus features comparably lower peak fields. For the measurement of THz power, on the other hand, basically all generated power is detected for all three waist diameters. The aperture of the power meter used is 20 mm, significantly larger than the diameter of the imaged THz beam, ensuring capture of all THz radiation.

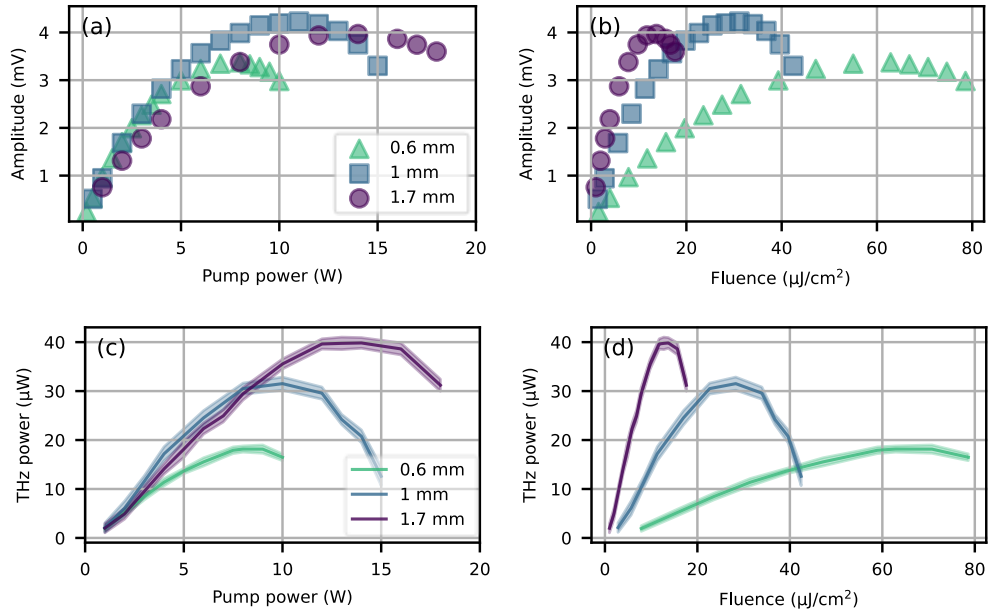

Fig. 6. (a) Peak-to-peak value of the EOS trace versus pump power. (b) Peak-to-peak value of the EOS versus pump fluence. (c) THz power versus pump power. (d) THz power versus pump fluence. All measurement points are acquired with a 10 $V_{pp}$ unipolar square waveshape bias voltage with a 50% duty cycle. EOS data is recorded with low-noise lock-in detection, which is modulated at 997 Hz via the bias voltage. THz average power is measured with the THz power meter, modulated with the same bias voltage at 18 Hz. The shaded areas in (c, d) represent the standard deviation of THz power measurements over 128 averages. The legend indicates the excitation beam diameter.

For different beam diameter of 0.6 mm, 1 mm, and 1.7 mm, the pump power reaches correspondingly 8.5 W, 10 W, and 14 W before rollover (Fig. 6(a, c)). These optical power levels correspondingly lead to THz powers of 18 µW, 31 µW and 40 µW. At the largest beam diameter of 1.7 mm, the LAE safely handles up to 18 W of average power. If the pump fluence is considered instead of pump average power (Fig. 6(b, d)), it is shown that at the smallest beam diameter (0.6 mm), more than 60 µJ/cm² of pump fluence could be applied to the LAE before reaching saturation, showing the main difficulty in scaling both efficiency and high average power simultaneously in these structures. For the 1 mm and 1.7 mm beam diameter, the maximum optical fluence before saturation was approximately 28 µJ/cm² and 13.5 µJ/cm², respectively. Increasing the beam diameter on the emitter enables us to apply higher pump power on the LAE before saturation, simultaneously leading to an increase in the THz power.

There are two main mechanisms that result in saturation of THz emission. One is screening of the bias field by the generated THz field at high fluences. This typically results in constant THz emission above a certain fluence and does not strongly depend on the size of the excitation beam waist [22]. Heating of the substrate, on the other hand, results in an increase of the effective mass of carriers either by nonparabolicity of the Γ-valley or side valley transfer. In particular the latter is expected to result in a pronounced reduction of THz emission as the effective mass in the L-valley $m_L = 1.9\ m_0$ is 30 times larger than $m_\Gamma = 0.063\ m_0$. The observation that in our experiment the THz emission decreases at higher power and the fact that the onset does not occur at a particular fluence indicates that heating by average power is the main issue here. Note that excitation at 516 nm (2.42 eV) high above the GaAs band gap of 1.42 eV in this regard is not ideal as the access energy contributes to the heating and photogenerated carriers are transferred with higher probability to the L-valley. A potential approach to tackle this issue would be to develop an LAE utilizing another photoconductive material adapted to direct excitation. Recently, we demonstrated photoconductive receivers based on ErAs:InAlGaAs for this wavelength [32]. Adapting this material to an LAE structure should be straightforward and with additional heatsinking, this should already result in improvements in conversion efficiency.

It should be noted that less than 20% of the incident radiation is absorbed by the LAE, due to the metallization structure covering 75% of the emitter surface and Fresnel loss at the air-GaAs interface [22]. In principle, this can be circumvented with lens arrays that would focus only on the active areas [19].

*3.2 LAE saturation behavior: optical modulation, constant electrical bias*

We aim here to disentangle the effects of the single pulse energy and the average power. In order to do this, the optical beam is modulated by an optical chopper with a 50% duty cycle, ensuring that the single pulse energy remains unchanged while the average power is halved. The THz power is measured using the THz power meter while the excitation beam is modulated at 18 Hz and a 10 V DC voltage is applied for bias voltage.

Fig. 7(a) shows the measured THz power as a function of pump power, where solid lines and dash lines correspond to optical and bias voltage modulations, respectively (om and bm in the figure caption). For different beam sizes (0.6 mm, 1 mm and 1.7 mm), the pump power reaches 5 W, 7 W and 9 W (after an optical chopper with 50% duty cycle) before reaching rollover. These pump power levels result in THz powers of 20 µW, 40 µW and 65 µW, respectively. For each spot size, a lower pump power on the LAE leads to higher THz power. Fig. 7(b), shows the measured THz power versus pump fluence, where solid lines and dash lines correspond to optical and bias voltage modulations, respectively. At the smallest 0.6 mm spot size, by decreasing the average power more than 80 µJ/cm² of pump fluence could be applied to the LAE before reaching saturation. For the 1 mm and 1.7 mm beam diameter, the maximum optical fluence before saturation is approximately 40 µJ/cm² and more than 18 µJ/cm², respectively. When fluences of the optical modulation (solid line) are compared to the fluences of the bias modulation (dashed line), it becomes evident that the saturation fluence is shifted to higher fluences with a lower pump average power.

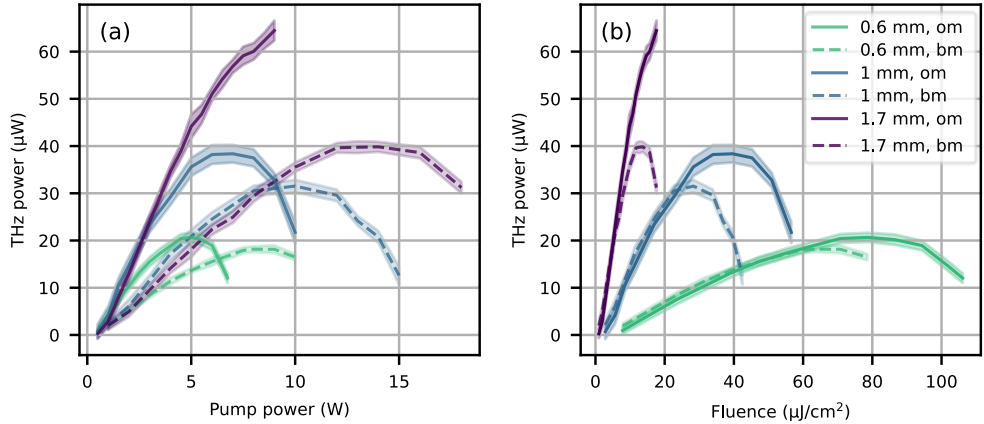

Fig. 7. (a) THz power versus pump power. (b) THz power versus pump fluence. Solid lines corresponded to THz power measurements with optical modulation (om) and dashed lines correspond THz power measurement with bias modulation (bm), in section 3.1. The THz power is acquired with optically modulation of green pump via optical chopper at 18 Hz and 50% duty cycle. For all the measurement points, the bias voltage is fixed at 10 V DC. The shaded areas present the standard deviation of the THz power measurement over 128 averages. om: modulation of optical pump beam via mechanical chopper with 50% of duty cycle. bm: modulation of the bias voltage by apply unipolar square waveshape with 50% of duty cycle. The legend indicates the excitation beam diameter.

By decreasing the pump power at constant pulse energy, the THz average power is increased for all spot sizes. This contributes to increasing the conversion efficiency as a result of more moderate thermal effects. These measurements confirm that the saturation effect in section 3.1 is mainly due to average power rather than pulse energy. In high repetition rate and high average power excitation, the pump average power is the primary obstacle for THz power scaling. We did not increase the beam size beyond 1.7 mm in diameter to avoid clipping of the laser beam on the active area of the LAE. In fact, a 1.7 mm is $1/e^2$ diameter Gaussian beam contains only 86.5% of the total optical power and a significantly larger beam would start suffering from clipping and loss in the 1 cm$^2$ square structure.

In this case, highest measured THz power of 65 µW is achieved with 9 W of excitation power (17.6 µJ/cm² excitation fluence), with a NIR-to-THz conversion efficiency of $3.6 \times 10^{-5}$. The efficiency is still lower than in previously reported oscillator-based excitation of an LAE [21]. This can be attributed to the excitation wavelength of 516 nm, which is significantly shorter than the optimal 800 nm wavelength, as explained above.

### 3.3 LAE limitations: optical versus electrical load

Following the exploration of the influence of fluence and average power on the most efficient regime of operation of the LAE, we aim here to disentangle the relative influences of electrical and optical heat load. The maximum optical power that could be safely applied to the emitter was 18 W at a 1.7 mm beam diameter (Fig. 6(a)), and the experiments in this section are conducted at this specific point. All measurements are carried out over a 78 s timeframe, corresponding to 155 traces of THz pulses.

#### 3.3.1. Impact of electrical power

To investigate the impact of electrical power, the optical power is kept constant at 18 W and is directed onto the emitter without any optical modulation. A unipolar rectangular bias voltage of 10 V$_{pp}$ with adjustable duty cycles is applied to the LAE. We mentioned above that a bipolar bias voltage offers better performance – however, this does not allow to directly study the influence of electrical load in similar conditions to an optical chopper where the excitation is on-off. Therefore, we chose here the unipolar case, which applies a power load comparable to the optical chopper. The acceleration field for the photoinduced carriers is proportional to the bias voltage. With a unipolar bias voltage, the acceleration field is maintained constant for all duty cycles, while the dissipated electrical power load varies with the duty cycle. In Fig. 8 bias voltage modulation is schematically illustrated with constant bias voltage and variable electrical power. A burst of unipolar rectangular wave shape with duty cycle of 50% and 20% are shown in Fig. 8(a) and Fig. 8(b), respectively. For each duty cycle the amplitude of bias voltage (correspondingly acceleration field) is constant, but the average electrical load is decreasing as duty cycle is reduced. The red dashed line shows the relative average electrical power. The optical pulse trains remain unmodulated with fixed pulse energy and average power for all duty cycle of bias voltages (Fig. 8(c)).

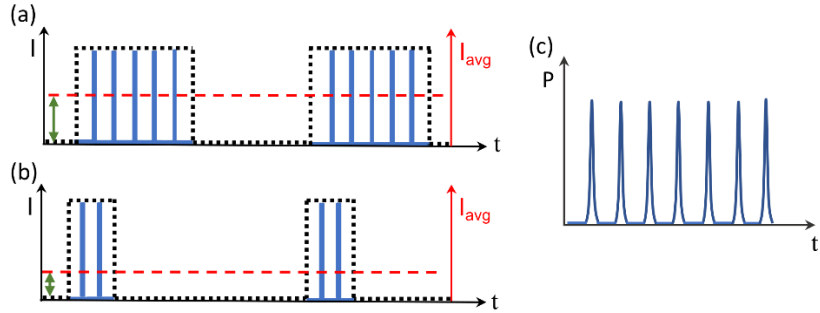

Fig. 8. Schematic illustration of bias voltage modulation at constant voltage value for different duty cycles. (a) modulation with 50% duty cycle. (b) modulation with 20% duty cycle. The red dashed line indicates relative average electrical power. (c) optical power remains unmodulated with 18 W of average power.

The duty cycle of the bias voltage varies in the ranging from 90% to 10% at 997 Hz. In this context, a duty cycle of 10% signifies that the voltage is on for 10% of each cycle. Fig. 9(a) displays the THz pulses in the time domain for each corresponding duty cycle. In Fig. 9(b), the THz signal in the frequency domain is presented. As one can see, the spectral characteristic does not change considerably with electrical power load. In Fig. 9(c) the peak-to-peak value of the EOS signals in the time domain is extracted for each corresponding duty cycle. In Fig. 9(d), the peak-to-peak value of the EOS signals from Fig. 9(c) are normalized to the peak-to-peak value of the 90% duty cycle to characterize how much the THz electric field increases as the duty cycle decreases to 10%. In all the plots, the color bar corresponds to the duty cycle of the bias voltage, as indicated in the legend of Fig. 9(a).

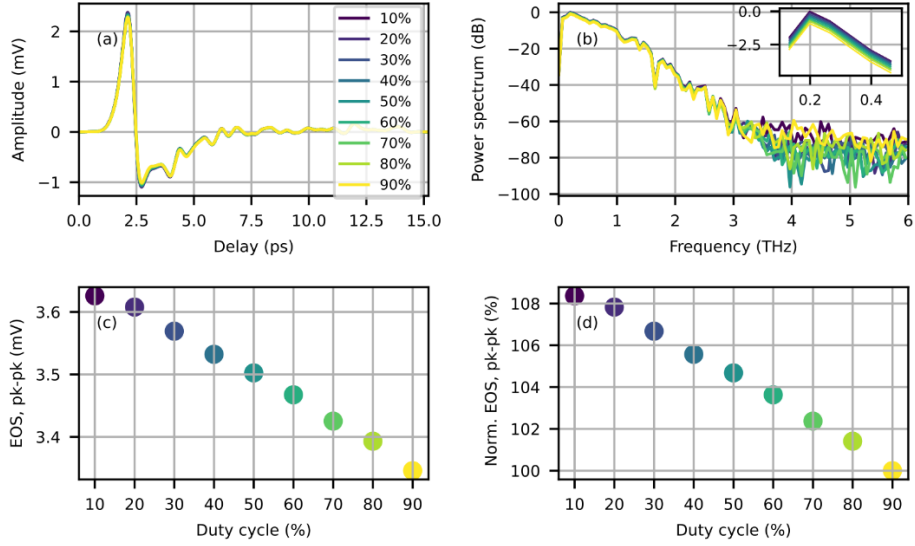

Fig. 9. THz generation as a function of duty cycle of the bias voltage. (a) THz trace in the time domain with various duty cycles of rectangular unipolar bias voltage, the y-axis representing the voltage from the balanced photodetector, which is proportional to the THz electric field. Legend shows the corresponding duty cycle of the bias voltage. (b) Corresponding power spectra of the traces in (a). (c) peak-to-peak value of EOS signals in the time domain. (d) normalized peak-to-peak value of EOS signals corresponding to (c). For all of measurement data, the pump power is set at 18 W and 1.7 mm spot diameter on LAE and each THz traces are averaged over 155 traces (78 s). Inset: zoom into data in frequency range between 0.1 THz to 0.5 THz, axes have the same unit as original plot.

By decreasing the duty cycle of the bias voltage from 90% to 10%, the peak-to-peak value of EOS signal increases by 8% (Fig. 9(d)), the corresponding single pulse THz E-field increases 8%, and the THz pulse energy $E_p$ increases by a factor of $1.08^2 = 1.17$. These measurements show an increase in the efficiency by a factor of 1.17 indicating the influence of electrical thermal load.

Notice that decreasing the duty cycle of the bias voltage by the factor of 9 also leads a reduction of effective repetition rate $f_{eff}$ (number of the pulses per second) of THz pulses with the same factor. The THz average power relatively decreased based on effective repetition rate to $P_{avg} = E_p \times f_{eff} \propto 1.17/9 \approx 13\%$. However, decreasing the duty cycle from 90% to 10% does not show a significant decrease in DR, while the number of averaged THz pulses in each trace and THz power are decreased simultaneously.

### 3.3.2 Impact of optical power

In a similar approach to the one used in the previous section we now investigate the impact of optical power load. The bias voltage is kept constant at 10 V DC (maintaining the same acceleration field). The excitation beam is modulated via an optical chopper with an adjustable duty cycle, which covers a range of 50% to 10% at 997 Hz. In this context, a duty cycle of 10% signifies that the optical beam goes through the chopper for 10% of each cycle and for the rest of the cycle it is blocked. In Fig. 10 optical power modulation is schematically depicted with a constant pump pulse energy and variable average power. Pump pulse burst with duty cycles of 50% and 20% are shown in Fig. 10(a) and Fig. 10(b), respectively. For each duty cycle, the single pulse energy remains constant, while the average power decreases as the duty cycle is reduced. The red dashed line indicates the relative average optical power. The bias voltage remains a constant DC value for all duty cycles during optical modulation. The pump power before the chopper is set to 18 W. Then, by varying the duty cycle of chopper from 50% to

10%, the impinging power on the LAE can decrease from 9 W down to 1.8 W, respectively, while the single pulse energy remains constant at ~200 nJ for all duty cycles.

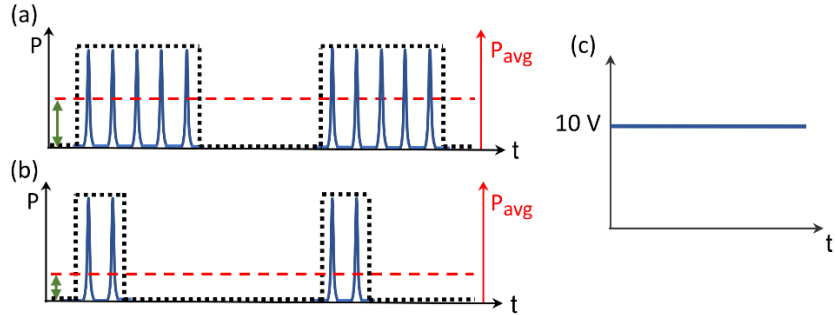

Fig. 10. Schematic illustration of optical modulation at constant optical pulse energy for different duty cycles. (a) modulation with 50% duty cycle. (b) modulation with 20% duty cycle. The red dashed line indicates relative average optical power. (c) bias voltage remains constant at 10 V DC.

Fig. 11(a) presents THz traces in the time domain for each duty cycle of the optical chopper, along with its corresponding average power as indicated in the legend. In Fig. 11(b), the THz signal in the frequency domain is illustrated. As it can be observed, the dynamic range slightly decreases with a smaller duty cycle, attributable to the decrease in THz power. Fig. 11(c) shows the peak-to-peak value of the EOS signals in the time domain versus each duty cycle of the optical chopper. In Fig. 11(d), the peak-to-peak values of the EOS signals from Fig. 11(c) are normalized to the signal at a 50% duty cycle to show how much the THz electric field increases as the duty cycle or average power decreases. In all the plots, the color bar corresponds to the duty cycle - or corresponding average power as indicated in the legend of Fig. 11(a).

By decreasing the duty cycle of the optical chopper from 50% to 10%, the peak-to-peak value of the EOS signal increases by 37% (Fig. 11(d)), the corresponding single pulse THz E-field increases by 37%, and the THz pulse energy $E_p$ increases $1.37^2 = 1.88$. Since the pump pulse energy and bias field are fixed for all duty cycles, the NIR-to-THz conversion efficiency is expected to be constant. However, this measurement shows an increase in the efficiency by 1.88, indicating the influence of optical thermal load.

Decreasing the duty cycle of optical chopper from 50% to 10%, also leads to a reduction of effective repetition rate $f_{eff}$ (number of the pulses per second) of THz pulses by a factor of 5. This leads to a relative decrease in THz average power to 37% ($P_{avg} = E_p \times f_{eff} \propto 1.88/5 \approx 37\%$). However, decreasing the duty cycle from 50% to 10% does not show a considerable decrease in DR, while the number of averaged THz pulses in each trace and THz power are decreased simultaneously.

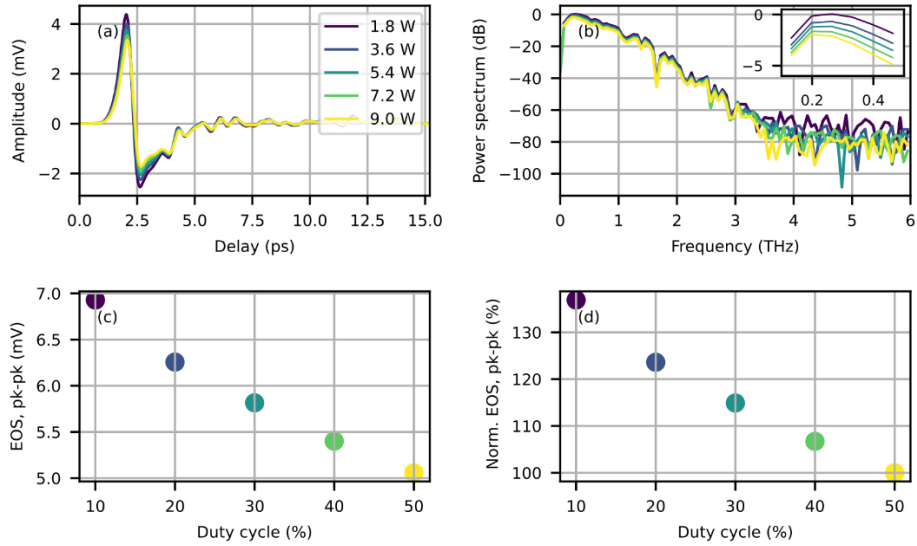

Fig. 11. THz generation as a function of duty cycle of the optical chopper. (a) THz trace in time domain with various duty cycles of the optical chopper, with the y-axis representing the voltage from the balanced photodetector, which is proportional to the THz electric field. Legend shows the incident optical average power on the LAE for each corresponding duty cycle of the optical chopper. (b) Corresponding power spectra of the traces in (a). (c) peak-to-peak value of time domain THz traces shown in (a). (d) normalized peak-to-peak value of EOS signals corresponding to (c). For all the measurement data, the pump power is set at 18 W before optical chopping and a 1.7 mm spot diameter is used on LAE. Each THz trace is an average over 155 traces (78 s). Inset: zoom into data in frequency range between 0.1 THz to 0.5 THz, axes have the same unit as original plot.

This result confirms that the saturation effect in the THz conversion efficiency by increasing the duty cycle is due to the optical pump power, demonstrating that the main limiting effect in this high average power, high repetition rate regime are thermal effects. This points to future improvements of the setup by improving cooling even further, i.e for example using back cooling in reflection geometry [33]. We cannot currently disentangle and distinguish the effect of heating caused either by the high photon energy or by the high average power alone. In future studies, we will compare this by using adapted materials for direct excitation with a 1030 nm laser, which can also be excited at 515 nm, by comparing the effects of the two wavelengths.

### 3.3.3. Discussion of electrical versus optical load

The heating of the LAE, which is the main source of limited emission THz power in our study, is caused by two effects. One is the direct optical absorption of the radiation, the other is Ohmic dissipation. The latter has a negligible effect compared to the photocurrent. Overall, the absorbed optical power, going up to ~3 W at 18 W incident power is larger than the Ohmic dissipated power of 0.9 W (93 mA average current at 10 V). Considering the two schemes of modulation it is important to note that full power direct optical heating is present for the electric modulation. Thus, reducing the electrical duty cycle reduces the total dissipated power only a comparable small fraction. Optical modulation with a lower duty cycle, on the other hand, reduces both the direct optical heating and the dissipated photocurrent, thus the reduction of heat is more significant in this case. There is probably also a geometrical effect. The optical power is absorbed in a few µm layer at the surface due to the large absorption coefficient for green light. The Ohmic dissipation, on the other hand, occurs in a larger volume. Here the depth is determined by the electric field distribution, around 5 µm for the 10 µm electrode gaps. Thus, the temperature increase by the same amount of dissipated energy is larger for optical absorption than for Ohmic dissipation in the LAE. Another candidate for further reducing thermal effects are bias-free photoconductive emitters. This type of emitter utilizes built-in electric fields to drive the photocarriers [34]. By designing bias-free LAEs, Ohmic heating can be eliminated. This could be a promising candidate for future studies.

## 4. Conclusion and outlook

We explore THz generation using a microstructured LAE with a home-built high-average-power thin-disk oscillator at 91 MHz repetition rate, frequency doubled to the green. The power available for LAE excitation is up to 18 W. We systematically study the limiting factors for THz conversion efficiency in this high repetition rate and high average power excitation regime. We find that thermal load due to the optical power has significantly more impact on LAE saturation than pulse energy in MHz/low-energy excitation regime. The impact of optical power and electrical power on the LAE conversion efficiency is investigated individually clearly indicating that thermal effects are the main limiting factor to scale efficiency. In optimized conditions with 9 W of pump power (after an optical chopper with duty cycle of 50%) and 1.7 mm spot diameter, we reach THz power of 65 µW with a NIR-to-THz conversion efficiency of $3.6 \times 10^{-5}$, spectrum span up to 4 THz with spectral DR about 85 dB. The efficiency is still lower than in previously reported oscillator-based excitation of an LAE [21] by close to one orders of magnitude; our study however allowed us to identify the causes of this reduced efficiency: mainly the excitation wavelength and thermal effects. We are confident that by tackling these 1) with the design of new photoconductive material for the right wavelength and 2) heatsinking the structure for example with substrate-removal and diamond contacting – efficiencies could be comparable to the lower power cases and very high THz average powers could be obtained. We note also that only a small fraction of power impingent on the structure is absorbed due to the metallization. Future mitigation strategies could include the use of microlenses to focus the power only where photoconductive material is present.

Our results represent the first step towards high-average-power THz source with a high repetition rate and high dynamic range based on LAE. The availability of high-average-power ultrafast sources at MHz or even GHz repetition rates at 1030 nm are a strong motivation for designing LAEs for these wavelengths, potentially enabling to approach high-power level THz sources at high repetition rates.

**Funding.** Ruhr-Universität Bochum (Open Access Publication Funds); Deutsche Forschungsgemeinschaft (287022738 TRR 196, 390677874); Ministerium für Kultur und Wissenschaft des Landes Nordrhein-Westfalen (terahertz.NRW).

**Acknowledgments.** The research work presented in this paper was funded by the German Research Foundation ("Deutsche Forschungsgemeinschaft") (DFG) under Project-ID 287022738 TRR 196 for Project M01. Funded by the Deutsche Forschungsgemeinschaft (DFG, German Research Foundation) under Germany's Excellence Strategy – EXC-2033 – Projektnummer 390677874 - Resolv. The project "terahertz.NRW" receives funding from the programme "Netzwerke 2021", an initiative of the Ministry of Culture and Science of the State of Northrhine Westphalia. The sole responsibility for the content of this publication lies with the authors.

**Disclosures.** The authors declare no conflicts of interest.

**Data availability.** Data underlying the results presented in this paper are available in Ref.[35].